\documentclass[aps,prl,twocolumn,superscriptaddress,showpacs,floatfix]{revtex4}

\usepackage[ansinew]{inputenc}
\usepackage{epsfig}
\usepackage{graphicx}
\usepackage{color}
\usepackage{bm}
\usepackage{multirow}
\usepackage{mathrsfs}
\usepackage{revsymb}
\usepackage{amssymb}
\usepackage{amsmath}
\usepackage{slashed}
\usepackage{longtable}
\usepackage{dcolumn}
\usepackage{upgreek}
\usepackage{textcomp}
\usepackage{endnotes}

\newcommand{\addrmpik}{Max--Planck--Institut f\"{u}r Kernphysik, Saupfercheckweg~1, 69117 Heidelberg, Germany}
\newcommand{\addrmainz}{Institut f\"ur Physik, Johannes Gutenberg-Universit\"at, 55099 Mainz, Germany}

\newcommand{\addrgsi}{GSI Helmholtzzentrum für Schwerionenforschung GmbH, Planckstra\ss{}e 1, 64291 Darmstadt, Germany}
\newcommand{\addrPet}{Department of Physics, St. Petersburg State University, Oulianovskaya 1, Petrodvorets, St. Petersburg 198504, Russia}
\newcommand{\addrdresden}{Institut für Theoretische Physik, Technische Universität Dresden, Mommsenstraße 13, 01062 Dresden, Germany}

\begin{document}

\title{\emph{g} factor of lithiumlike silicon $^{28}$Si$^{11+}$}

\author{A.~Wagner}
\affiliation{\addrmpik}
\author{S.~Sturm}
\affiliation{\addrmpik}
\affiliation{\addrmainz}
\author{F.~K\"{o}hler}
\affiliation{\addrmpik}
\affiliation{\addrgsi}
\author{D.~A.~Glazov}
\affiliation{\addrPet}
\affiliation{\addrdresden}
\author{A.~V.~Volotka}
\affiliation{\addrPet}
\affiliation{\addrdresden}
\author{G.~Plunien}
\affiliation{\addrdresden}
\author{W.~Quint}
\affiliation{\addrgsi}
\author{G.~Werth}
\affiliation{\addrmainz}
\author{V.~M.~Shabaev}
\affiliation{\addrPet}
\author{K.~Blaum}
\affiliation{\addrmpik}

\date{\today}

\begin{abstract}
The \emph{g} factor of lithiumlike $^{28}$Si$^{11+}$ has been measured in a triple-Penning trap with a relative uncertainty of $1.1\cdot10^{-9}$ to be $g_{\rm exp}=2.000\,889\,889\,9(21)$. The theoretical prediction for this value was calculated to be $g_{\rm th}=2.000\,889\,909(51)$ improving the accuracy to $2.5\cdot10^{-8}$ due to the first rigorous evaluation of the two-photon exchange correction. The measured value is in excellent agreement with the state-of-the-art theoretical prediction and yields the most stringent test of bound-state QED for the \emph{g} factor of the $1s^22s$ state and the relativistic many-electron calculations in a magnetic field.
\end{abstract}

\maketitle

Many electron atomic systems have been subject to a vast number of experimental and theoretical investigations. Among them lithiumlike three-electron systems are of particular interest since their properties can be calculated with high accuracy and comparison with experiments provide stringent tests of the theory \cite{kozhedub2007qed}. This holds particularly for bound-state quantum electrodynamic (BS-QED) effects which manifest themselves e.g., in the magnetic moment of the bound electrons, expressed by the \emph{g} factor. The many-electron contributions to the \emph{g} factor are of purely relativistic origin, which necessitates their consideration within the \emph{ab initio} QED approach (see, e.g. \cite{shabaev2002two} and references therein). This can be illustrated by analyzing the contributions of the negative energy Dirac states, which are of the same order of magnitude as those of the positive-energy states \cite{PhysRevA.47.961, lithium, indelicato2003correlation, glazov:04:pra}. The fully relativistic treatment to all orders in $\alpha Z$ of the interelectronic-interaction effects is possible only within a field theoretical approach, namely QED. For the \emph{g} factor of lithiumlike ions the main theoretical uncertainty for all values of \emph{Z} is determined by the interelectronic-interaction correction. While the one-photon exchange diagrams were calculated in Ref.~\cite{lithium}, the two-photon exchange contribution is known only in the leading order in $\alpha Z$ \cite{glazov:04:pra}. The rigorous evaluation of the two-photon exchange contribution to the \emph{g} factor has been obtained in this Letter for the first time and resulted in an improvement of the theoretical uncertainty for the interelectronic-interaction correction by a factor of three. The total theoretical accuracy for the \emph{g} factor of $^{28}$Si$^{11+}$ has been improved by almost a factor of two.\\
Experiments to determine the \emph{g} factor of lithiumlike ions can use methods developed for the investigation of hydrogenlike ions, which have resulted in very high accuracy of the \emph{g} factor of the single bound electron \cite{Haeff2000, Verdu2004, sturm2011g} and represent presently the most accurate test of BS-QED. Accordingly, the investigation of the \emph{g} factor of lithium-like ions serves as an excellent test for QED calculations of many-electron systems.\\
\begin{figure}[b,t]
	\centering
		\includegraphics[width=0.5\textwidth]{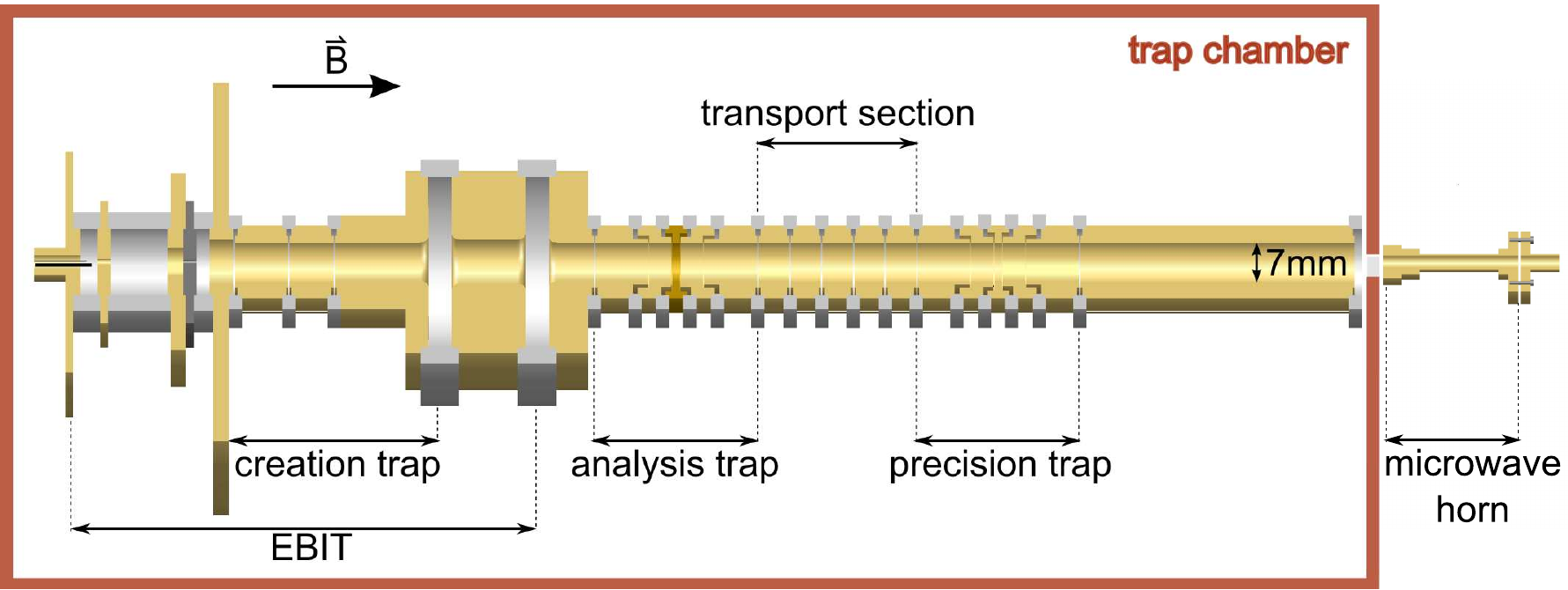}
	\caption{(Color online) The trap arrangement consists of an electron beam ion trap (EBIT) for the creation of highly-charged ions, an analysis trap where a ferromagnetic ring electrode is used to produce a magnetic inhomogeneity and a precision trap with a very homogeneous magnetic field. In order to induce spin flips, microwaves are irradiated through a quartz glass window into the trap chamber.}
	\label{fig1}
\end{figure}
The BS-QED contributions increase with the nuclear charge $Z$. Therefore, it is advantageous to measure the \emph{g} factor of heavier elements, provided that the experimental and theoretical uncertainties remain approximately independent of $Z$. However, the uncertainty due to the nuclear size effect increases strongly with \emph{Z} and sets limits on the possibility to probe QED effects. Considering the nuclear size contributions for hydrogen- and lithiumlike ions it was found in Ref.~\cite{lithium} that due to the similar behaviour of the $1s$ and $2s$ wave functions, their ratio is very stable with respect to variations of nuclear models. This means that the uncertainty of the nuclear effects can be significantly reduced with a combination of the \emph{g} factors of hydrogen- and lithiumlike ions, providing an additional motivation for the investigations of lithiumlike ions on the same level of accuracy as hydrogenlike ones. Moreover, many-electron systems will be of importance for the possibility to determine the fine structure constant $\alpha$ independently from QED with experiments on heavy highly charged ions \cite{shabaev:06:prl}.

This Letter reports on the \emph{g} factor measurement of lithiumlike silicon $^{28}$Si$^{11+}$ and the comparison with the theoretical value. It represents the most precise \emph{g} factor determination of a three-electron system to date. Compared to the measurements on $^{6,7}$Li \cite{boklen1966high} using the atomic beam magnetic resonance method and $^9$Be$^+$ \cite{PhysRevLett.50.628} by laser-induced fluorescence, the experimental uncertainty is reduced by two orders of magnitude. Moreover, the sensitivity to relativistic effects is increased due to the higher nuclear charge $Z$. In our experiment we follow the procedure for our \emph{g} factor determination of hydrogenlike silicon $^{28}$Si$^{13+}$ \cite{sturm2011g,schabinger2012experimental}.

A single ion is created and confined in a triple-Penning trap system of 7\,mm inner diameter (see Fig.\,\ref{fig1}), which is located in a strong magnetic field of $B_0=3.76$\,T. For the creation of the highly-charged ions the principle of an electron beam ion trap (EBIT) is employed \cite{Alonso2006}. Via electron bombardment neutral atoms and singly ionized ions are released from a target. The latter are confined in a three-electrode Penning trap (``creation trap''), and are consecutively ionized by an electron beam of 4\,keV to produce higher charge states. Afterwards the ion cloud, containing many species with different charge-to-mass ratios, is adiabatically transported to a five-electrode Penning trap, the so-called ``precision trap'', where the ions experience a strong homogeneous magnetic field. Here unwanted species and charge states are removed from the trap by excitation of their respective oscillation amplitudes. The number of the selected ions is finally reduced to a single one, whose axial motion is then resistively cooled to the ambient temperature of 4.2\,K.

Experimentally the \emph{g} factor can be determined by measuring the Larmor frequency $\nu_L$ of the bound electron in a magnetic field $B_0$
\begin{equation}
\nu_L=\frac{g}{4\pi}\frac{e}{m_e}B_0
\end{equation}
with $e$ and $m_e$ being the charge and mass of the electron, respectively. To determine the magnetic field strength $B_0$, the ion with mass $M$ and charge $q$ itself is used as a magnetic probe by measuring its free cyclotron frequency $\nu_c=q\,B_0/(2\pi M)$. Accordingly, the \emph{g} factor measurement can be reduced to a measurement of the frequency ratio $\Gamma\equiv\nu_L/\nu_c$:
\begin{equation}
g=2\frac{\nu_L}{\nu_c}\frac{q}{e}\frac{m_e}{M}=2\Gamma\frac{q}{e}\frac{m_e}{M}.
\label{Eq1}
\end{equation}
The motion of the ion within the Penning trap is a superposition of three independent harmonic eigenmotions, which can be measured non-destructively to high precision: the modified cyclotron frequency $\nu_+\simeq$\,23\,MHz, the axial frequency $\nu_z\simeq$\,687\,kHz and the magnetron drift frequency $\nu_-\simeq$\,10\,kHz. The free cyclotron frequency is obtained by applying the Brown-Gabrielse invariance theorem $\nu_c^2=\nu_+^2+\nu_z^2+\nu_-^2$ \cite{gabrielse2009sideband}, where frequency shifts arising from misalignments between trap and magnetic field cancel to first order. All information on the ion's motion is determined through the axial mode. By adjusting the ring voltage, the axial oscillation is brought into resonance with a superconducting tank-circuit, which is attached to one end-cap electrode and has a quality factor of 950. The oscillating ion induces mirror charges in the trap electrodes and the resulting current, of the order of fA, leads to a voltage drop across the tank-circuit's resistor. This voltage drop is amplified by a low-noise amplifier and afterwards analysed with a fast Fourier transformation. The ion's axial frequency can be detected as a minimum, a so-called ``dip'', in the thermal noise spectrum of the resonator, which appears at the oscillation frequency of the ion. The precise value of the axial frequency is deduced from a fit to the data \cite{schabinger2012experimental}.

The radial modes are coupled to the axial mode by a radio frequency field at the sideband frequencies $\nu_+-\nu_z$ and $\nu_z+\nu_-$, respectively. This causes the axial dip to split into two minima (``double-dip'') and allows for the measurement of the reduced cyclotron and magnetron frequency, respectively \cite{sturm2010g, schabinger2009creation}.
\begin{figure}[t]
	\centering
		\includegraphics[width=0.48\textwidth]{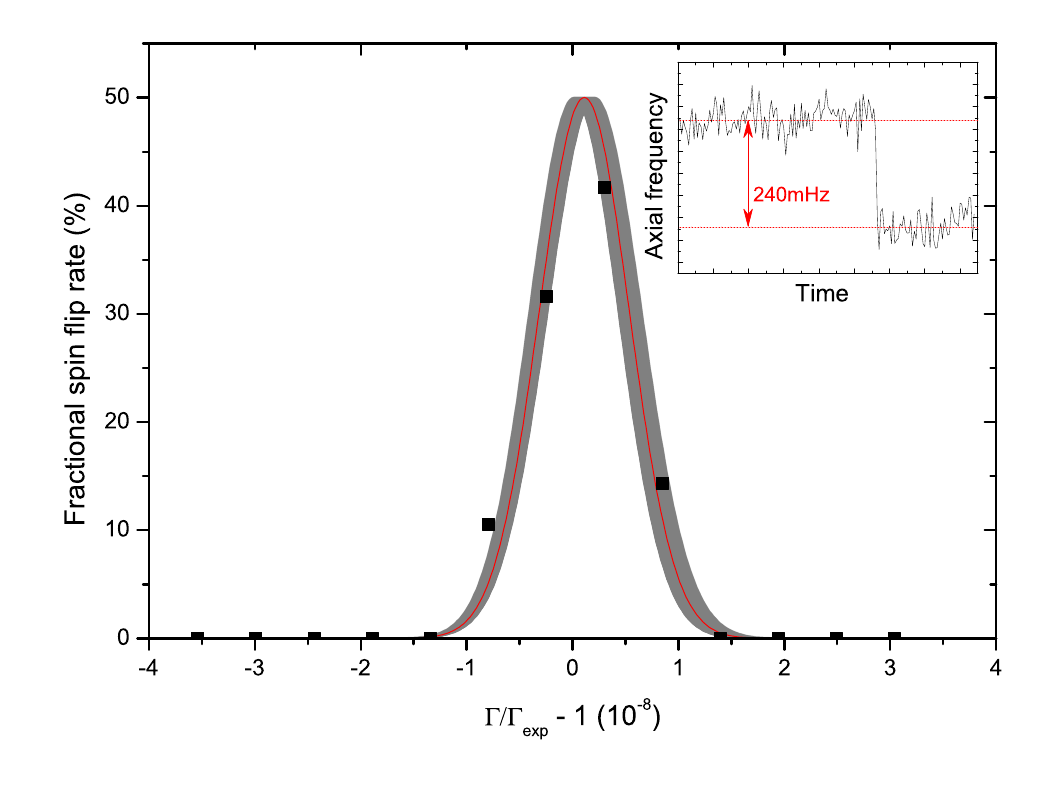}
	\caption{(Color online) Spin-flip resonance of a single lithiumlike silicon ion. A Gaussian lineshape (solid line) was fit to the data. The dark grey area is the confidence interval of the Maximum-Likelihood fit. 
	Due to a high microwave power the resonance is saturated. The inset shows a discrete spin flip, which is monitored by a jump of 240\,mHz within the absolute axial frequency of about 411.8\,kHz.}
	\label{fig2}
\end{figure}

In order to measure $\nu_L$, microwaves close to the expected Larmor frequency, of nominally 105\,GHz, are irradiated into the trap to induce a spin flip. The spin direction can be determined in the third trap, the ``analysis trap'', by employing the continuous Stern-Gerlach effect \cite{werth2002continuous}. This trap is identical to the precision trap except for the material of the ring electrode. Here, a ferromagnetic electrode is used to deliberately induce a magnetic inhomogeneity, which results in a quadratic dependence of the magnetic field on the axial coordinate: $B_\textnormal{z}=B_0+B_2z^2+...$ . Within this so-called magnetic bottle the spin orientation is coupled to the axial frequency. The ion, due to its magnetic moment $\mu$, experiences an additional force $\vec{F}=\vec{\mu}\nabla\vec{B}=2\mu_{\textnormal{z}} B_2\hat{z}$ which, depending on the spin orientation, either adds or subtracts to the electric trapping force. Accordingly, a spin flip manifests as a small change of the axial oscillation frequency of the ion $\Delta\nu_z=(g\mu_B B_2)/(4\pi^2M\nu_z)$ with $\mu_B=e\hbar/(2m_e)$ being the Bohr magneton. For our experimental parameters ($B_2$=10\,mT/mm$^2$, $M$=28\,u, $\nu_z$=411.8\,kHz), the frequency jump amounts to $\Delta\nu_z$=240\,mHz. Detecting this tiny change requires very stable trapping conditions, especially a voltage source with a stability of $\delta U/U\le2.5\cdot10^{-7}$ and constant motional amplitudes of the radial modes. Indeed, as can be seen in the inset of Fig.\,\ref{fig2}, the frequency stability obtained is extremely high and the two spin states can be unambiguously distinguished.

The \emph{g} factor measurement cycle starts with the ion located in the analysis trap. After determination of the electron's spin direction, the ion is adiabatically transferred to the precision trap where microwaves are irradiated to induce a spin flip and, simultaneously, the oscillation frequencies are measured. After transport back to the analysis trap, it is determined whether the spin orientation has changed. This measurement cycle is repeated several hundred times while the microwave frequency is varied. A resonance of the spin flip probability obtained as a function of the respective frequency ratio $\Gamma$ is shown in Fig.\,\ref{fig2}.

As a result of the necessity to observe at least one spin flip in the analysis trap, the required measurement time depends on the spin flip probability in the analysis trap. For our setup, the theoretical lineshape yields a spin flip probability of the order of 30\,$\%$, similar as in the hydrogenlike system \cite{sturm2011g}. However, due to a misalignment of the microwave coupling into the trap in addition to a displacement of the ion inside the strong magnetic inhomogeneity of the analysis trap, large magnetron sidebands were observed and caused the maximum spin flip probability in the analysis trap to be only $\sim$1\,$\%$. This did not affect the \emph{g} factor resonances measured in the precision trap and thus the final \emph{g} factor value, but limited the statistics.\\
\begin{figure}[t]
	\centering
	\includegraphics[width=0.4\textwidth]{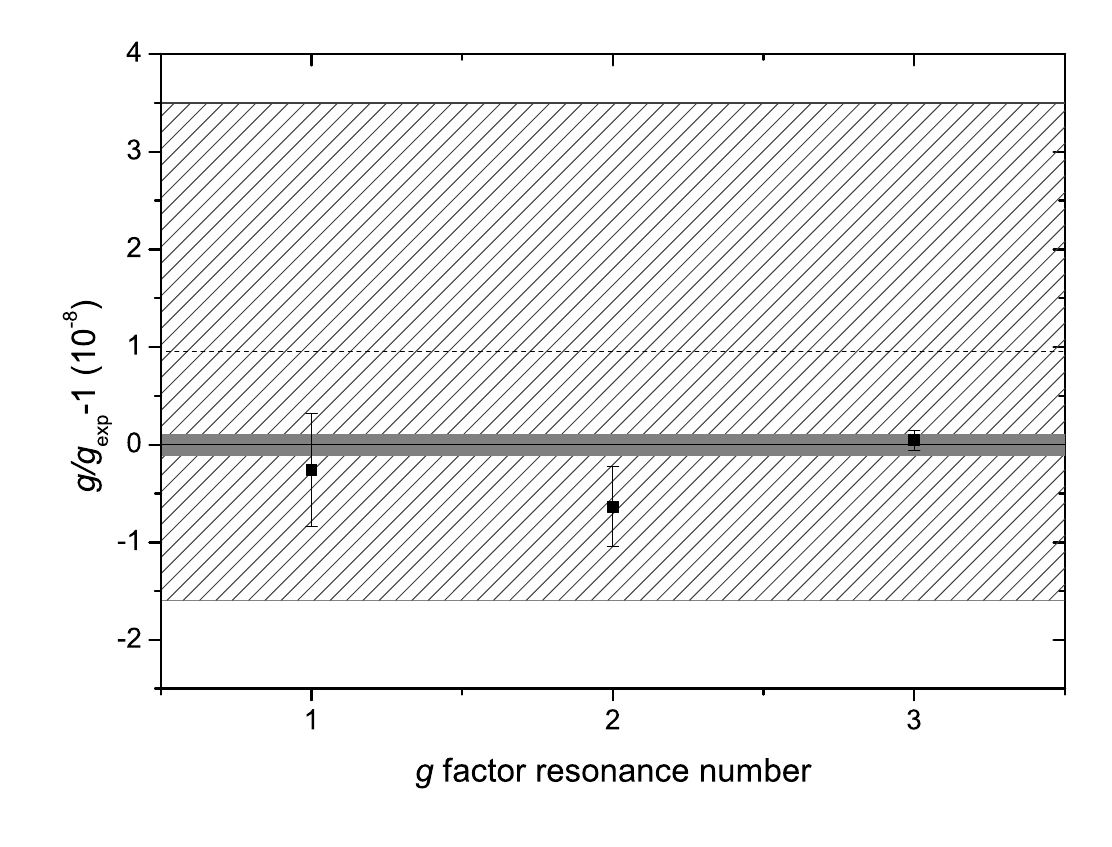}
	\caption{Comparison of experimental and theoretical \emph{g} factor for $^{28}$Si$^{11+}$. Three spin flip resonances have been recorded which are marked by the black points. Overall, the microwave power was varied by a factor of $\sim$50. The solid line is the resulting weighted average with the corresponding experimental uncertainty in dark grey. The dashed line denotes the theoretical value and the hatched area its uncertainty.}
	\label{fig3}
\end{figure}
We recorded three \emph{g} factor resonances (see Fig.\,\ref{fig3}) with different microwave power. Although all three resonances were saturated, no power dependence of the extracted \emph{g} factor value was observed. The frequency ratios $\Gamma_\textnormal{exp,(i=1,2,3)}$ are obtained by fitting a Gaussian lineshape to the data points employing the Maximum-Likelihood method to avoid a binning of the data. The weighted average for the three resonances is $\Gamma_\textnormal{exp}'=4637.318\,949(4)$. This value has to be corrected for systematic shifts. The main shift arises from the interaction of the ion with its induced mirror charges in the trap electrodes \cite{PhysRevA.40.6308, PhysRevA.64.023403}. Due to the low free cyclotron frequency of $^{28}$Si$^{11+}$ the retarded contribution to the mirror charge effect can be neglected. For our trap it was calculated that $\delta\nu_{\textnormal{c}}/\nu_{\textnormal{c}}=6.58(33)\cdot 10^{-10}$ \cite{Sturm}. This uncertainty is below the statistical uncertainty and thus does not affect the final \emph{g} factor value. For the corrected frequency ratio we obtain $\Gamma_\textnormal{exp}=4637.318\,946(4)$. The dominant systematic error arises from the determination of the axial frequency, which is deduced from a fit of a well-known lineshape to the axial dip. However, the characteristics of the resonator and the detection system are required as input parameters for this lineshape. They can only be determined to a limited precision and therefore result in a systematic error for the \emph{g}-factor of $\delta g/g=2.6\cdot 10^{-10}$ \cite{schabinger2012experimental}. Relativistic shifts are of the order of $10^{-13}$ or below and can be neglected.

To calculate the \emph{g} factor from Eq.\,(\ref{Eq1}) we take $m_\textnormal{e}=5.485\,799\,094\,6\,(22)\cdot 10^{-4}$\,u for the electron mass \cite{Mohr2012} and $M(^{28}$Si$^{11+}$)=27.970\,894\,575\,81(66)\,u from \cite{PhysRevLett.100.093002}, corrected for the masses of the eleven missing electrons and their binding energies \cite{martin1983energy}. With these values we derive the final \emph{g} factor to
\begin{equation}
g_{\rm exp}=2.000\,889\,889\,9(19)(5)(8).
\end{equation}
Here, the three error bars are the statistical and systematic uncertainties as well as the uncertainty due to the electron mass, respectively. The total experimental precision is limited by the statistical uncertainty due to the low spin flip probability in the analysis trap and the correspondingly required long measurement time.

The theoretical contributions to the \emph{g} factor of lithiumlike ions can be split into one-electron and many-electron parts. The one-electron terms are determined by Feynman diagrams similar to those for hydrogenlike ions, while the many-electron effects are described by two- and three-electron Feynman diagrams and define the main difference in the calculations. The contributions to the theoretical value of the ground-state \emph{g} factor of lithiumlike silicon are presented in Table \ref{tab:g-th}. Details of the calculations of most of the relevant contributions can be found in Refs.~\cite{lithium,yan:2002:prl,yerokhin:04:pra,glazov:04:pra,pachucki:05:pra} and references therein.
\begin{figure}[t,b]
\centering
\includegraphics[width=0.4\textwidth]{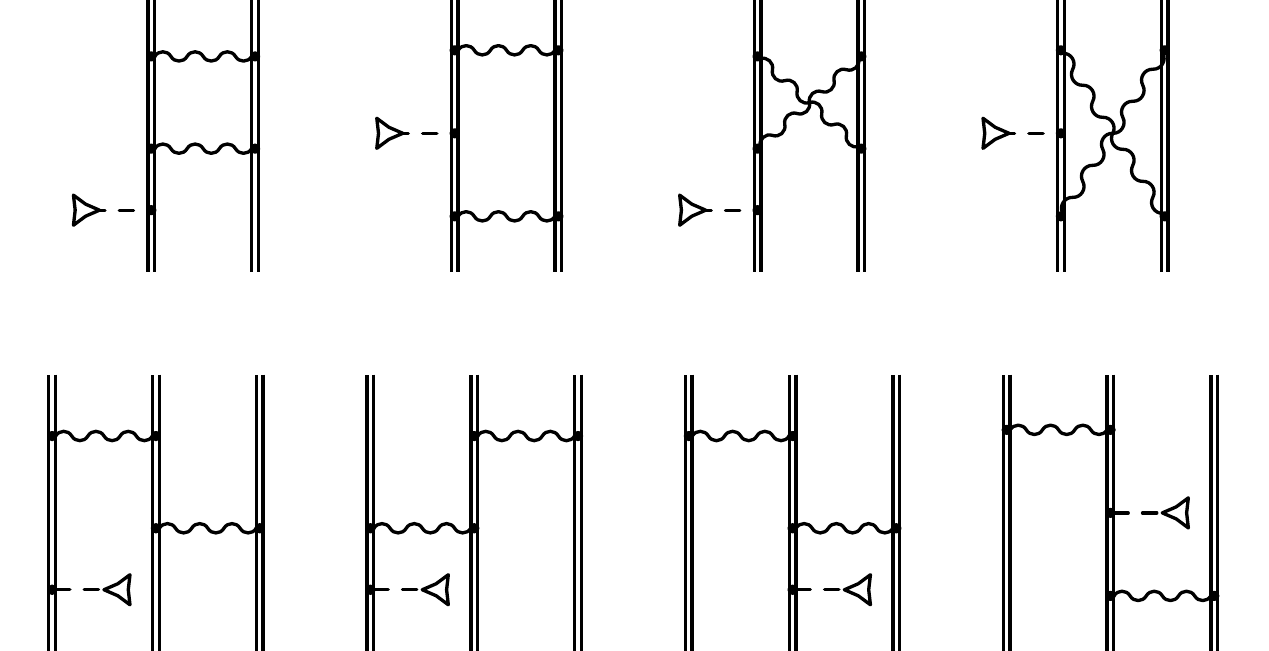}
\caption{Feynman diagrams representing the two-photon exchange correction to the \emph{g} factor. The wavy line indicates the photon propagator and the double line the electron propagators in the Coulomb field of the nucleus. The dashed line terminated with the triangle denotes the interaction with the constant magnetic field.}
\label{fig4}
\end{figure}
The uncertainty of the theoretical value
\begin{equation}
g_{\rm th}=2.000\,889\,909(51)
\end{equation}
is defined by the screened QED and interelectronic-interaction contributions. The screened QED correction was evaluated to the leading order $\alpha (\alpha Z)^2 / Z$ in Ref.~\cite{glazov:04:pra}. Rigorous evaluation of the two-electron self-energy and vacuum-polarization contributions to the \emph{g} factor of lithiumlike ions has been performed so far only for high nuclear charges \cite{volotka:09:prl,glazov:10:pra}. The one-photon exchange diagrams, which represent the interelectronic-interaction correction of first order in $1/Z$, were evaluated in Ref.~\cite{lithium}. Evaluation of the two-photon exchange diagrams in an external magnetic field (see Fig.\,\ref{fig4}), corresponding to the $1/Z^2$-term, remained a challenge for theory until recently. It has been conquered for the case of the hyperfine splitting in Ref.~\cite{volotka:12:prl}, and in this Letter we present the first corresponding result for the \emph{g} factor of lithiumlike silicon to be $-0.000\,006\,876(1)$. Thereby, the uncertainty of the total interelectronic-interaction correction has been improved by more than a factor of three in comparison with previous study \cite{glazov:04:pra}, from 0.000\,314\,903(74) to 0.000\,314\,801(22). The interelectronic-interaction contributions to the third and higher orders in $1/Z$ have been calculated by employing the large-scale configuration-interaction Dirac-Fock-Sturm method \cite{glazov:04:pra}. For the \emph{g} factor the interelectronic-interaction effects, regardless of the order in $1/Z$, are of pure relativistic origin. In particular, the contribution of the negative-energy Dirac continuum is not additionally suppressed by the factor $(\alpha Z)^2$. As a result, the contribution of the negative-energy states to the $1/Z$-term amounts to $-58$\,\% with respect to the total value, compared to 158\,\% from the positive-energy states. For the $1/Z^2$-term $-223$\,\% arises from the negative- and $323$\,\% from the positive-energy contribution. This demonstrates the special importance of rigorous QED treatment of the interelectronic-interaction in the case of \emph{g} factor even for light ions. Further improvement of the theoretical accuracy can be achieved by rigorous calculation of the screened QED contribution and the $1/Z^3$-term of the interelectronic interaction.\\
\begin{table}
\caption{Individual theoretical contributions to the ground-state \emph{g} factor of lithiumlike silicon $^{28}$Si$^{11+}$.\label{tab:g-th}}
\begin{tabular}{lr@{}l}
\hline
\hline
Dirac value (point nucleus)         &    1.&998 254 751      \\
Finite nuclear size                 &    0.&000 000 003      \\
Interelectronic interaction         &      &                 \\
  $\sim 1/Z$                        &    0.&000 321 592      \\
  $\sim 1/Z^2$                      & $-$0.&000 006 876 (1)  \\
  $\sim 1/Z^3$ and higher orders    &    0.&000 000 085 (22) \\
QED                                 &      &                 \\
  $\sim \alpha$                     &    0.&002 324 044 (3)  \\
  $\sim \alpha^2$ and higher orders & $-$0.&000 003 517 (1)  \\
Screened QED                        & $-$0.&000 000 212 (46) \\
Nuclear recoil                      &    0.&000 000 039 (1)  \\
Total                               &    2.&000 889 909 (51) \\
\hline
\hline
\end{tabular}
\end{table}
The comparison between the experimental and theoretical \emph{g} factors confirms the relativistic many-electron effects at the level of $10^{-4}$ and, in particular, the two-photon exchange contribution to $1$\,\%. Further improvement of the theoretical uncertainty to match the experimental accuracy will directly improve these tests by more than one order of magnitude.

In summary, we have presented the first high-precision measurement of the \emph{g} factor of the lithiumlike silicon ion $^{28}$Si$^{11+}$ with a relative uncertainty of \mbox{$1.1\cdot 10^{-9}$}. Excellent agreement between experiment \mbox{$g_{\rm exp}=2.000\,889\,889\,9(21)$} and theory \mbox{$g_{\rm th}=2.000\,889\,909(51)$} provides the most stringent test of bound-state QED for the \emph{g} factor of the 2s state and the relativistic many-electron calculations in a magnetic field. An extension of these investigations to heavier hydrogen-, lithium-, and boronlike systems will provide access to the QED effects in the strongest electromagnetic fields available for the experimental study.\\
This work was supported by the Max-Planck Society, the Helmholtz Alliance HA216/EMMI, the EU (ERC Grant No. 290870 - MEFUCO) and by the IMPRS-QD. The work of D.A.G., A.V.V., G.P. and V.M.S. was supported in part by DFG (Grant No. VO 1707/1-2), GSI, RFBR (Grant Nos. 12-02-31803 and 10-02-00450), and by a grant of the President of the Russian Federation (Grant No. MK-3215.2011.2). D.A.G. acknowledges financial support by the FAIR -- Russia Research Center and by the ``Dynasty'' foundation.

\bibliography{Literaturverweise}

\end{document}